\documentclass[intlimits,twoside,a4paper]{article}

\usepackage{graphicx} 

\usepackage{placeins}
\usepackage{subcaption}
\usepackage{setspace}
\usepackage{placeins}  
\usepackage{float}     

\usepackage{fontenc}  
\usepackage[utf8]{inputenc}

\usepackage{epstopdf}
\usepackage{amsmath}

\usepackage[eqsecnum]{cmpj3} 

\issue{2025}{28}{4}{43701}
\doinumber{10.5488/CMP.28.43701}

\title[Properties of Co$_2$YPb Heuslers]%
{First principle study of electronic, magnetic and thermoelectric properties of Co$_2$YPb (Y = Tc, Ti, Zr and Hf) full Heusler: Application to embedded automotive systems}

\author[N. Saidi, A. Abbad, W. Benstaali, K. Bahnes]{
N. Saidi\orcid{0009-0004-5343-8572},
A. Abbad\orcid{0009-0006-1622-5564},
W. Benstaali\orcid{0000-0003-4634-6210}\thanks{Corresponding author: \email{ben\_wissam@yahoo.fr}},
K. Bahnes\orcid{0009-0007-3676-1126}
}
\address{
Laboratory of Technology and Solids Properties, Faculty of Sciences and Technology, BP227, Abdelhamid Ibn Badis University, 27000 Mostaganem, Algeria
}


\date{Received March 28, 2025, in final form August 12, 2025}

\begin{document}

\maketitle

\begin{abstract}
In this study, theoretical investigation on structural, electronic, magnetic, elastic and thermoelectric properties of the full Heusler Co$_2$YPb (Y = Tc, Ti, Zr and Hf) alloys have been performed within density functional theory~(DFT). The exchange and correlation potential is addressed using two approximations: the generalized gradient approximation (GGA) and the GGA augmented by the Tran--Blaha-modified Becke--Johnson (mBj-GGA) approximation, which provides a more accurate description of the energy band gap. The electronic and magnetic properties reveal that the full-Heusler alloys Co$_2$YPb (with Y = Tc, Ti, Zr, and Hf) display half-metallic ferromagnetic behavior. Furthermore, the elastic properties suggest that Co$_2$YPb are mechanically stable, with ductile characteristics. \mbox{$p$-type} full Heusler alloys exhibit positive Seebeck coefficients and high $ZT$ values, indicating good thermoelectric performance in terms of electrical and thermal conductivity.  This leads us to the conclusions that these compounds are very interesting in improving the performance of embedded automotive systems and can also be used in spintronic devices.
\printkeywords 
%
\end{abstract}

\section{Introduction}
\label{sec:intro}

Heusler compounds were discovered in 1903 by the German chemical engineer Friedrich Heusler~\cite{Wollmann2017}, who synthesized the compound Cu$_2$MnAl. He observed that this material exhibited a ferromagnetic~(FM) character, although none of its constituent atoms were intrinsically ferromagnetic. In 1934, an X-ray diffraction study by Bradley and Rodger revealed that the ferromagnetic phase of this type of alloy adopts an ordered structure at room temperature \cite{Bradley1934}. Since their discovery, these alloys have attracted increasing scientific interest due to their potential applications in spintronic devices such as tunnel magnetoresistance (TMR), giant magnetoresistance (GMR) \cite{Chadov2011,Hara2006}, magnetic random-access memories~(MRAM) and magnetic sensors \cite{Wolf2001,DeGroot1983}, thermoelectric applications, skyrmion and superconductivity \cite{Kawasaki2022}.

In 1983, half-metallic ferromagnetic (HMF) was first predicted in the half-Heusler alloys NiMnSb and PtMnSb by de Groot et al. \cite{DeGroot1983}, as well as in the full-Heusler alloy Co$_2$MnSn by K\"ubler et al. \cite{Kubler1983}. The first Heusler alloys studied crystallized in an L2$_1$ structure composed of four face-centered cubic (fcc) sublattices. It was later found that certain alloys take on a C1b structure, with one of the four sublattices remaining vacant.

Half-metallic magnets (HMMs) are characterized by exhibiting a metallic conductivity for one spin channel, while the opposite spin channel behaves semi-conductively with an energy gap at the Fermi level. This unique electronic structure results in a fully spin-polarized state at the Fermi level, where the spin-up density of states at the Fermi level (N$_{\uparrow}$$_{\mathrm{EF}}$) is non-zero, while the spin-down density (N$_{\downarrow}$$_{\mathrm{EF}}$) is zero \cite{Junxiang2023}. This spin polarization enables the possibility of a completely spin-polarized current, making HMMs ideal candidates for applications in spintronic and magnetoelectronic devices due to their enhanced performance and efficiency.

Heusler alloys, known for their half-metallic ferromagnetic properties, are considered promising candidates for spintronic applications due to their high spin polarization \cite{Kostenko2018}. However, their electronic and magnetic properties are mainly determined by the hybridization between elements X and Y and by the nature of the $sp$ element Z \cite{Ivanshin2009}. Heusler alloys are generally classified into two main categories: half-Heuslers, which follow the general formula XYZ, and full-Heuslers, which have the formula X$_2$YZ. In these formulas, X and Y represent transition metals while Z denotes a III--V main group element \cite{Marchenkov2023}.

The first compound crystallizes in a C1b-type face-centered cubic structure, belonging to space group F$\bar{4}$3m (\#216). The second compound can adopts two possible structures: the Hg$_2$CuTi type structure, which also belongs to the F$\bar{4}$3m (216) space group \cite{Graf2009}, and the Cu$_2$MnAl type structure, which falls under the Fm$\bar{3}$m (225) space group \cite{Salaheldeen2022}. As for the third compound, it crystallizes in a LiMgPdSn-type structure with space group F$\bar{4}$3m (216), exhibiting three distinct crystal configurations (named type-1, type-2 and type-3) \cite{Benatmane2020}.
These structures are characterized by the interpenetration of four face-centered cubic (fcc) sublattices.

Among these compounds, in this study we choose to investigate the structural, electronic, magnetic, thermoelectric, and elastic properties of cobalt-based full-Heusler alloys. Co$_2$TiPb  has been  studied theoretically~\cite{Zitouni2020}. For the best of our knowledge, there are no expiremental data for the Co$_2$YPb (Y = Tc, Ti, Zr, and Hf).

\section{Computational details}
\label{sec:compdetails}

In this study, calculations were performed using the full-potential linearized augmented plane wave~(FP-LAPW) method, implemented in the WIEN2k code. Within the framework of density functional theory (DFT), the exchange and correlation energies were treated using the Perdew--Burke--Ernzerhof~(PBE) generalized gradient approximation (GGA) \cite{Perdew1998}, as well as the mBJ-GGA approximation, designed to optimize the energy bandgap values to better match the experimental results \cite{Tran2009, Tran2007}. We examined the structural stability between two structures: the Heusler Cu$_2$MnAl alloy and the inverse Heusler Hg$_2$CuTi alloy. In the interstitial region, the plane-wave cutoff value used was $RMT \times K_{\rm max} = 8$, where $RMT$ represents the smallest value of the muffin-tin sphere radii (MT), and $K_{\rm max}$ is the largest reciprocal lattice vector used in the plane wave. The maximum angular momentum value for the wave function expansion within the atomic spheres was limited to $l_{\rm max} = 10$. The charge density was expanded in Fourier series with $G_{\rm max} = 24$~(arb. units)$^{-1}$ \cite{Pagare2014}. The muffin-tin sphere radius ($RMT$) values for the Co, Tc, Ti, Zr, Hf, and Pb atoms used in our calculations are as follows: Co (2.4 a.u.), Tc (2.2 a.u.), Ti~(2.4~a.u.), Zr~(2.5~a.u.), Hf~(2.3~a.u.), and Pb (2.5 a.u.). These values are associated to the Co$_2$YPb compounds, where~Y is Tc, Ti, Zr, or Hf. The self-consistent calculations are considered converged when the total energy of the system stabilizes to 10$^{-4}$ Ry. To explore the magnetic behavior, both spin-polarized (for the FM state) and non-spin-polarized [for the non-magnetic (NM) state] calculations were performed. For the spin-polarized calculations, in the initialization part, we use the instructions (init-up then init-dn) to explore the magnetism and for non-magnetic calculation we use the instruction (init-nm). These initializations ensure a reliable convergence towards the correct magnetic groud state. The initialization mesh parameter used was taken from \cite{Houari2019}.

\section{Results and discussions}
\label{sec:results}

\subsection{Structural properties}
\label{subsec:struct}

The key to an access to  physical properties of full-Heusler alloys Co$_2$YPb (Y= Tc, Ti, Zr and Hf) is to first study their structural properties and find a stable state \cite{Monkhorst1976, Houari2019, Mentefa2021}. 

Total energies corresponding to various volumes were evaluated within the framework of the GGA approximation. The calculations were done for two types of structures: the ``Cu$_2$MnAl-type'' structure in space group Fm$\bar{3}$m, \#225, and the ``Hg$_2$CuTi-type'' inverse structure in space group F$\bar{4}$3m, \#216. The calculations were performed for both the ferromagnetic and non-magnetic states, the curves are represented using the Birch--Murnaghan equation \cite{Murnaghan1944}:
\begin{equation}
\label{eq:BM}
E(V) = E_0 + \frac{B_0 V}{B_0'(B_0'-1)} \left[ B_0' \left(1 - \frac{V_0}{V}\right) + \left(\frac{V_0}{V}\right)^{B_0'} - 1 \right],
\end{equation}
where $E_0$ represents the total energy at equilibrium when the temperature is at 0 K, $V_0$ is the volume of the unit cell at equilibrium, $B_0$ is the bulk modulus and $B'_0$ is the first derivative of the bulk modulus concerning pressure.

\begin{figure}[htbp]
	\centering
		\includegraphics[scale=0.7]{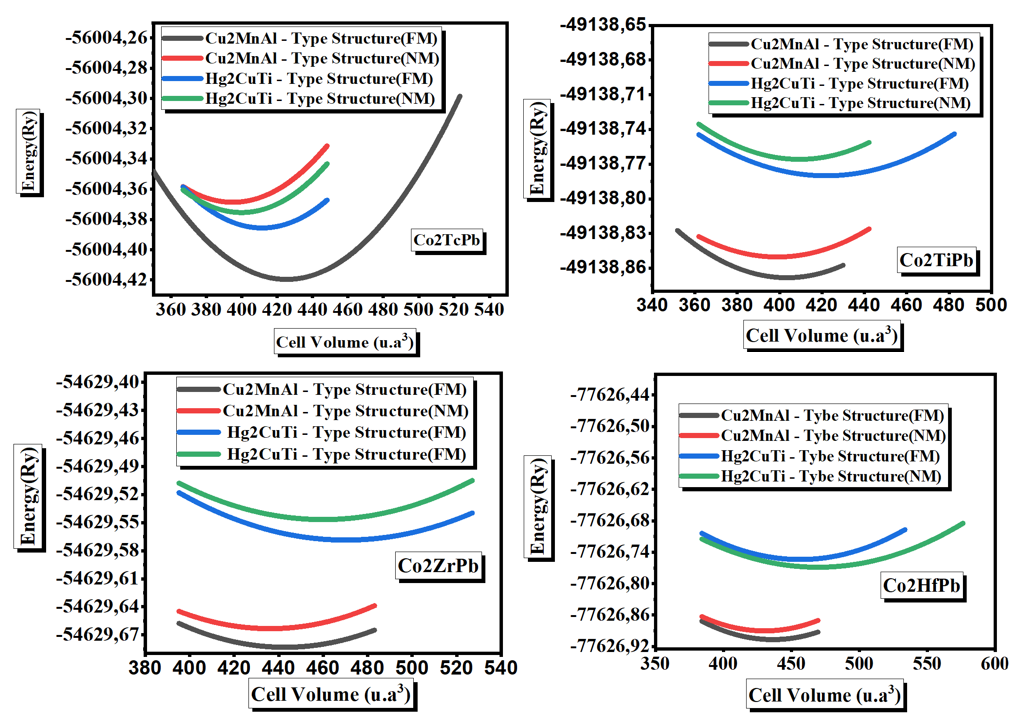}
	
	\caption{(Colour online) Variation of total energy evolution of the full Heusler Co$_2$YPb alloys (Y = Tc, Ti, Zr and Hf) as a function of unit cell volume, for Cu$_2$MnAl and Hg$_2$CuTi type structures, in FM and NM states.}
	\label{fig:figure1}
\end{figure}

Figure~\ref{fig:figure1} displays the energy variations of four Heusler alloys as a function of cell volume. It can be observed that the Heusler compounds, Co$_2$TcPb, Co$_2$TiPb, Co$_2$ZrPb and Co$_2$HfPb, converge towards the minimum energy in the ferromagnetic state of the Cu$_2$MnAl structure. This indicates that the most stable structure in which the studied alloys crystallize is the Cu$_2$MnAl structure. Therefore, the properties of these alloys were analyzed based on this structure.

To evaluate the stability of these compounds, we calculated their formation energy per atom at zero temperature. This value was obtained by using the formula \cite{Karimian2015} 
and is shown in the following equation:
\begin{equation}
\label{eq:formE}
\Delta H = E_{\text{formation}}^{\text{Co}_2\text{YPb}} = E_{\text{total}}^{\text{Co}_2\text{YPb}} - \big(2 E_{\text{Bulk}}^{\text{Co}} + E_{\text{Bulk}}^{\text{Y}} + E_{\text{Bulk}}^{\text{Pb}}\big).
\end{equation}
The formation energy represents the difference between the energy of a crystal $E_{\text{total}}^{\text{Co}_2\text{YPb}}$ and the sum of the energies of the elements that make it up ($E_{\text{Bulk}}^{\text{Co}}$, $E_{\text{Bulk}}^{\text{Y}}$ et $E_{\text{Bulk}}^{\text{Pb}}$). Negative values of formation energy are a clear sign of thermodynamic stability. Results in table~\ref{tab:01} reveal that all Co$_2$YPb (Y=Tc, Ti, Zr and~Hf) compounds have negative formation energy values, indicating that they are thermodynamically stable and can be synthesized experimentally.

\begin{table}[!htb]
\caption{Calculated equilibrium lattice parameters ($a_0$), bulk modulus ($B$), derivative of bulk modulus~($B'$), equilibrium volume ($V_0$), minimum energy ($Eg$), and formation energy ($\Delta H$) of the full-Heusler Co$_2$YPb (where Y = Tc, Ti, Zr and Hf) alloys with GGA.}
\label{tab:01}
\vspace{2ex}
\begin{center}
\begin{tabular}{lcccccc}
\hline
Alloys & $a_0$ (\AA{}) & $B$ (GPa) & $B'$ & $V_0$ ($r_{\textrm{Bohr}}^3$) & $Eg$ (Ry) & $\Delta H$ (Ry/atom) \\ 
\hline
Co$_2$TcPb & 6.22 & 156.8556 & 4.7322 & 407       & $-56004.41695$ & $-1.423$ \\
Co$_2$TiPb & 6.19 (6.20 \cite{Houari2019}) & 150.9835 & 4.7371 & 401.44255 & $-49138.86721$ & $-1.389$ \\
Co$_2$ZrPb & 6.38 & 141.4195 & 4.8992 & 439.1654  & $-54629.6830$  & $-1.439$ \\
Co$_2$HfPb & 6.35 & 147.2040 & 4.8551 & 433.3244  & $-77626.90610$ &  $-0.876$   \\ 
\hline
\end{tabular}
\end{center}
\end{table}

\subsection{Electronic properties}
\label{subsec:electronic}

The study of electronic properties aims to clarify the insulating, semiconducting, metallic or even semi-metallic nature of a solid, as well as the nature of the chemical interaction phenomena linking the different atoms that make it up. This study will give us a better understanding of the majority physical properties on a macroscopic scale, as these are strongly linked to the electronic structure. The electronic properties of the materials studied can be analyzed by calculating the band structure and the electronic density of state (DOS).

\subsubsection{Band structures}
\label{subsubsec:bands}

The spin-polarized electronic structure is an essential tool for describing the electronic properties of compounds. In this study, the spin-polarized electronic structure of the compound full Heusler Co$_2$YPb (Y = Tc, Ti, Zr and Hf) was analyzed at the equilibrium lattice parameters obtained along the directions of high symmetry in the first Brillouin zone, using the GGA and GGA approximations with the mBJ-GGA~\cite{Idriss2020}. 

\begin{figure}[h]
	\centering
	\begin{minipage}[b]{0.41\textwidth}
		\centering
		\includegraphics[width=\textwidth]{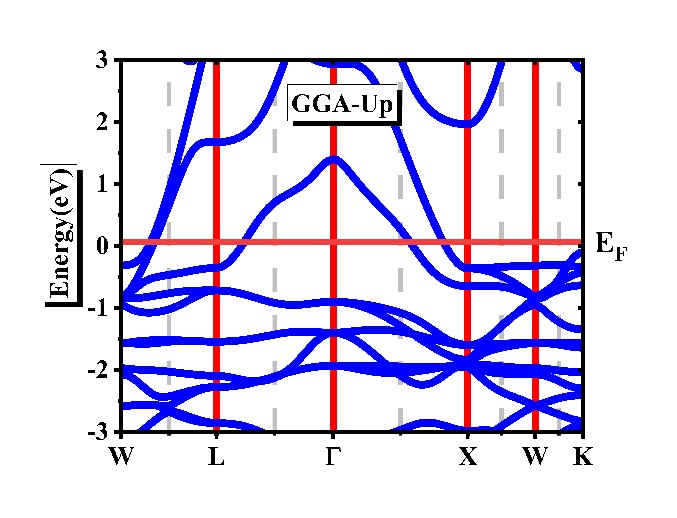}
	\end{minipage}
	\begin{minipage}[b]{0.41\textwidth}
		\centering
		\includegraphics[width=\textwidth]{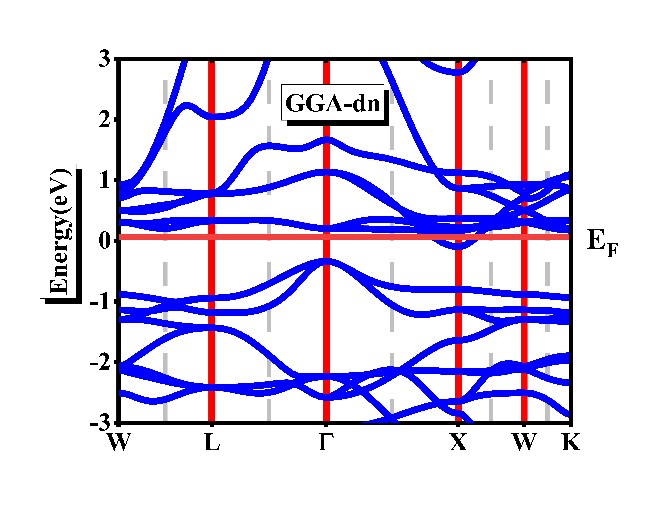}
	\end{minipage}
		\begin{minipage}[b]{0.41\textwidth}
		\centering
		\includegraphics[width=\textwidth]{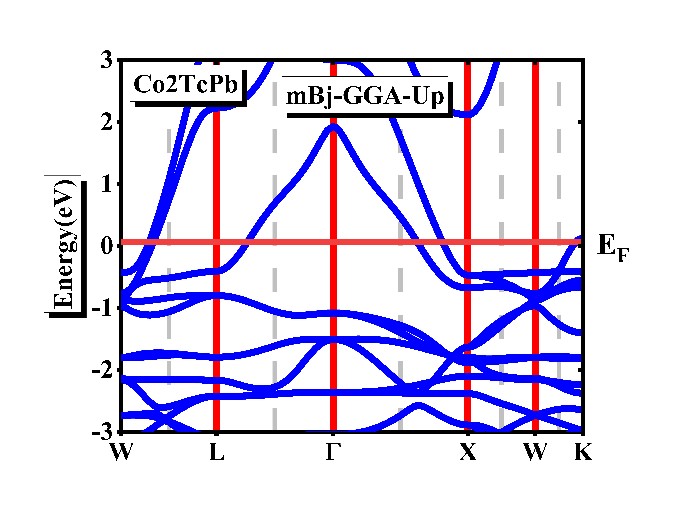}
	\end{minipage}
	\begin{minipage}[b]{0.41\textwidth}
		\centering
		\includegraphics[width=\textwidth]{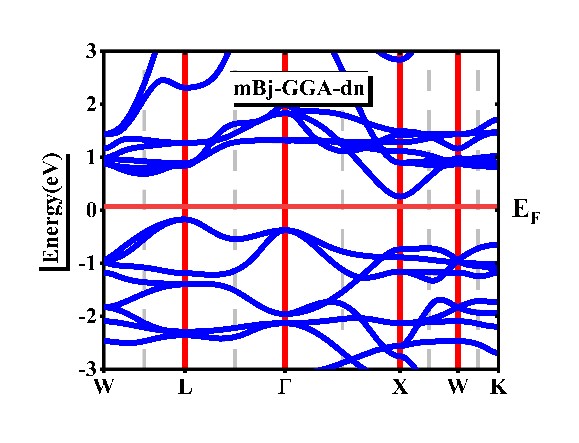}
	\end{minipage}
	\caption{(Colour online) The band structure of the full Co$_2$TcPb Heusler alloys calculated using the GGA and GGA-mBJ approximations to the equilibrium lattice parameter for spin-up and spin-down.}
	\label{fig:figure2}
\end{figure}

\begin{figure}[htbp]
    \centering
    \begin{minipage}[b]{0.45\textwidth}
        \centering
        \includegraphics[width=\textwidth]{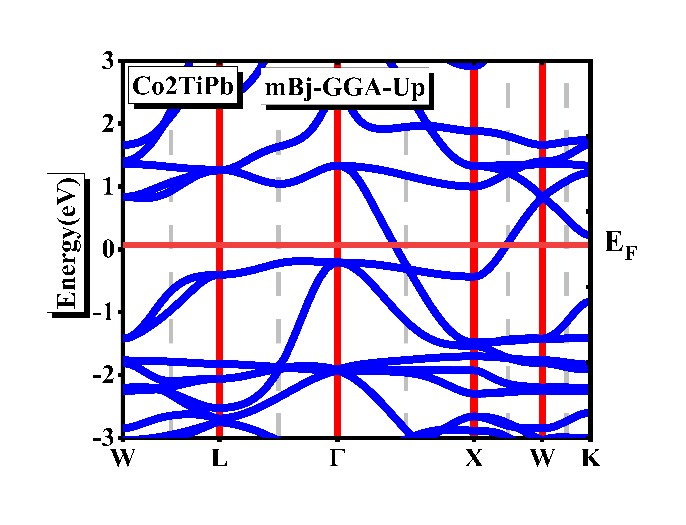}
    \end{minipage}
    \begin{minipage}[b]{0.45\textwidth}
        \centering
        \includegraphics[width=\textwidth]{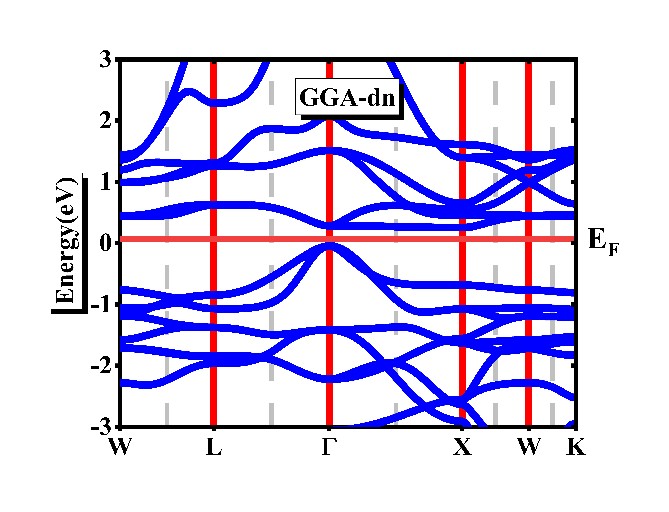}
    \end{minipage}
    \begin{minipage}[b]{0.45\textwidth}
        \centering
        \includegraphics[width=\textwidth]{assets/fig_3a}
    \end{minipage}
    \begin{minipage}[b]{0.45\textwidth}
        \centering
        \includegraphics[width=\textwidth]{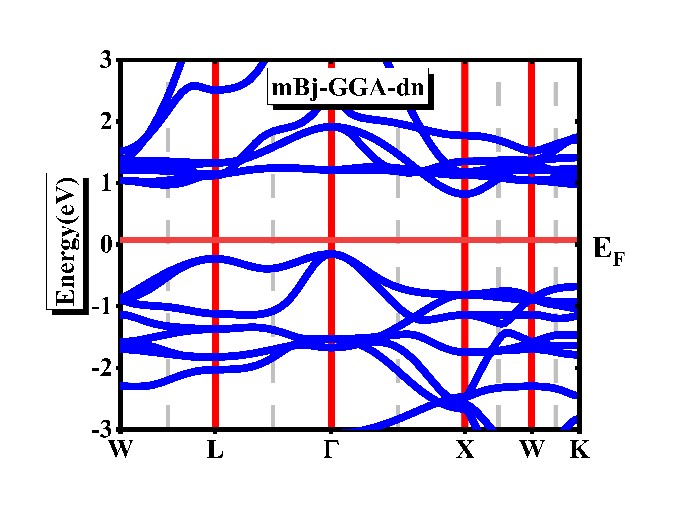}
    \end{minipage}
 \caption{(Colour online) The band structure of the full Co$_2$TiPb Heusler alloys calculated using the GGA and GGA-mBJ approximations to the equilibrium lattice parameter for spin-up and spin-down.}
    \label{fig:figure3}
\end{figure}

\begin{figure}[h]
    \centering
    \begin{minipage}[b]{0.41\textwidth}
        \centering
        \includegraphics[width=\textwidth]{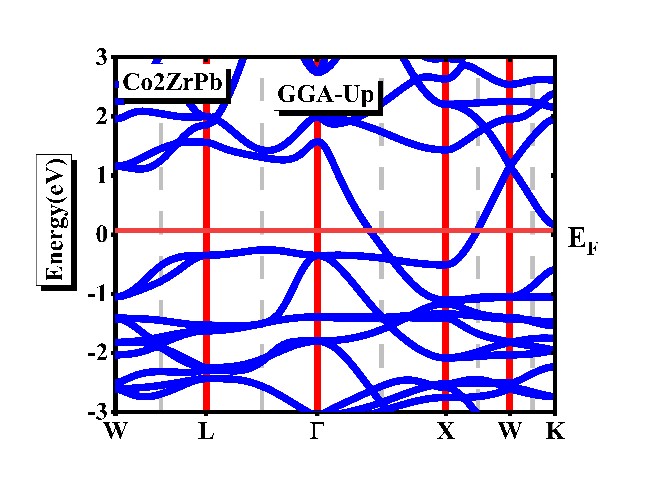}
    \end{minipage}
    \begin{minipage}[b]{0.41\textwidth}
        \centering
        \includegraphics[width=\textwidth]{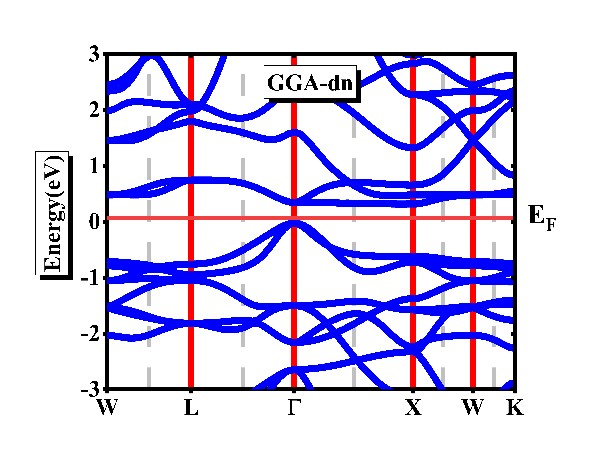}
    \end{minipage}
  \begin{minipage}[b]{0.41\textwidth}
        \centering
        \includegraphics[width=\textwidth]{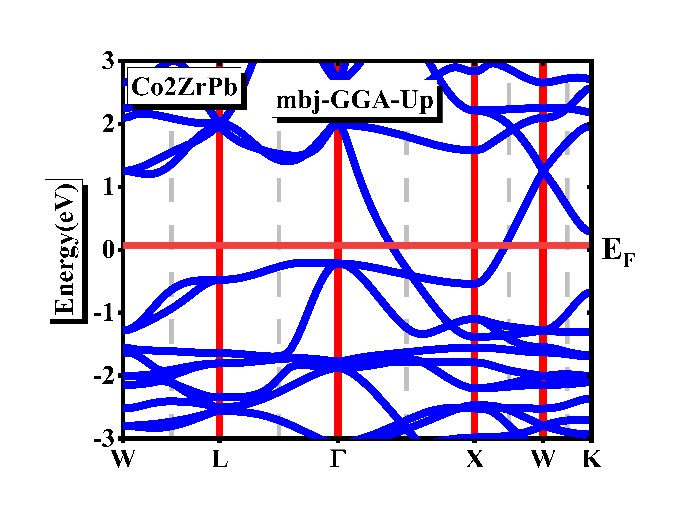}
    \end{minipage}
    \begin{minipage}[b]{0.41\textwidth}
        \centering
        \includegraphics[width=\textwidth]{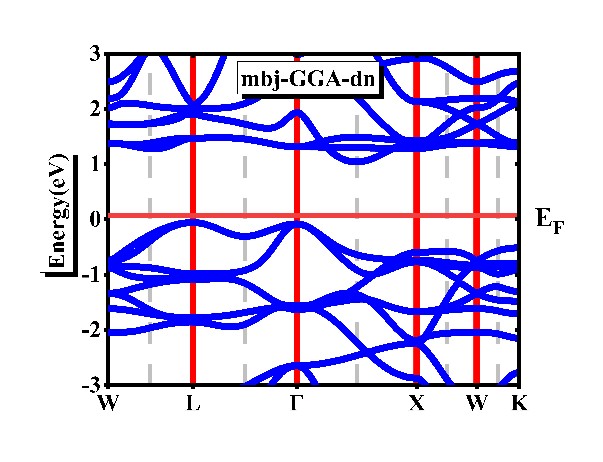}
    \end{minipage}
\caption{(Colour online) The band structure of the full Co$_2$ZrPb Heusler alloys calculated using the GGA and GGA-mBJ approximations to the equilibrium lattice parameter for spin-up and spin-down.}
    \label{fig:figure4}
\end{figure}

\begin{figure}[!h]
    \centering
    \begin{minipage}[b]{0.41\textwidth}
        \centering
        \includegraphics[width=\textwidth]{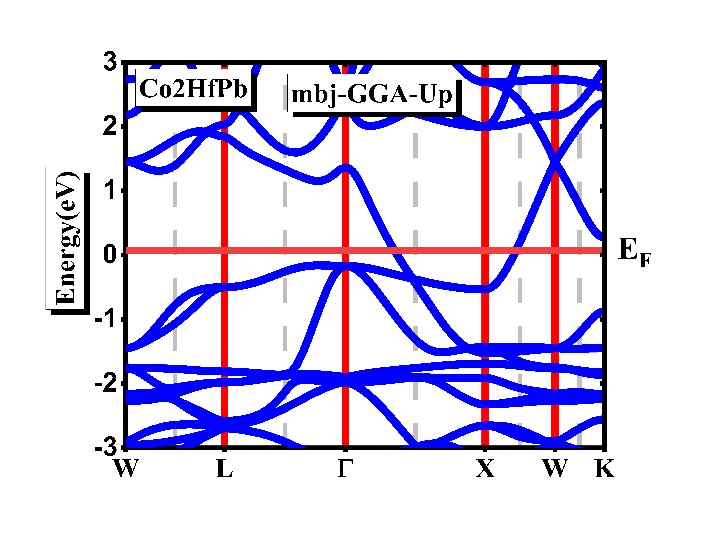}
    \end{minipage}
    \begin{minipage}[b]{0.41\textwidth}
        \centering
        \includegraphics[width=\textwidth]{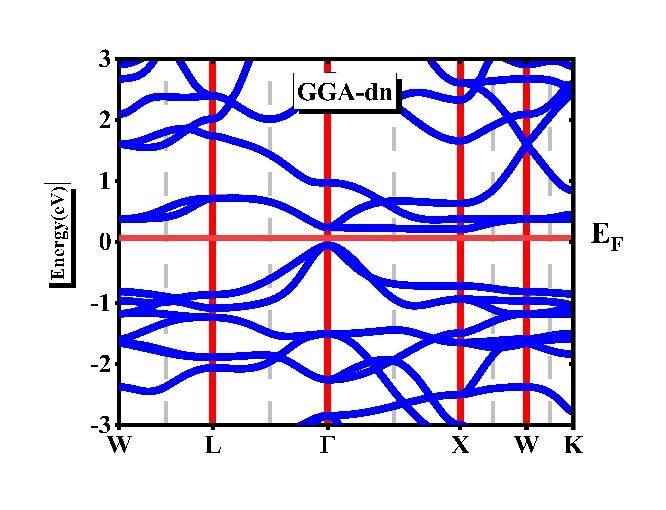}
    \end{minipage}
    \begin{minipage}[b]{0.41\textwidth}
        \centering
        \includegraphics[width=\textwidth]{assets/fig_5a}
    \end{minipage}
    \begin{minipage}[b]{0.41\textwidth}
        \centering
        \includegraphics[width=\textwidth]{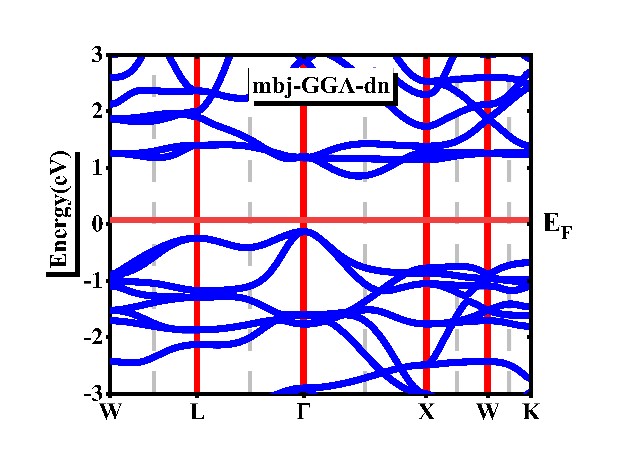}
    \end{minipage}
   \caption{(Colour online) The band structure of the full Co$_2$HfPb Heusler alloys calculated using the GGA and GGA-mBJ approximations to the equilibrium lattice parameter for spin-up and spin-down.}
    \label{fig:figure5}
\end{figure}

Based on the band structure depicted in {figures~\ref{fig:figure2},~\ref{fig:figure3},~\ref{fig:figure4} and~\ref{fig:figure5}}, we can gain valuable information regarding the electronic properties of the compounds Co$_2$YPb (Y = Tc, Ti, Zr, and Hf). On examination of the Fermi level, it is apparent that the compounds exhibit metallic properties in their spin-up state, but in the spin-down state, we notice the presence of an indirect band gap between the conduction and valence bands for Co$_2$TiPb (0.2 eV), Co$_2$ZrPb (0.3 eV) and Co$_2$HfPb (0.2 eV) but a metallic behavior for Co$_2$TcPb using the GGA approximation. And when applying the mBJ-GGA method, the obtained energy band gap is higher compared to the GGA approximation, with values of 0.5~eV, 0.9~eV, 1~eV, and 0.8~eV for Co$_2$TcPb, Co$_2$TiPb \cite{Houari2019}, Co$_2$ZrPb \cite{Houari2024}, and Co$_2$HfPb, respectively (see table~\ref{tab:02}). 

\begin{table}[!htb]
\caption{Calculated band gaps (in eV) for Co$_2$YPb (Y = Tc, Ti, Zr and Hf) using the GGA, and mBj-GGA approximations.}
\label{tab:02}
\vspace{2ex}
\begin{center}
\begin{tabular}{lcc}
\hline
Compounds & GGA (eV) & mBJ-GGA (eV) \\
\hline
Co$_2$TcPb & metal & 0.5 \\
Co$_2$TiPb & 0.2   & 0.9 \\
Co$_2$ZrPb & 0.3   & 1.0 \\
Co$_2$HfPb & 0.2   & 0.8 \\
\hline
\end{tabular}
\end{center}
\end{table}


\newpage

\newpage
\subsubsection{Density of states}
\label{subsubsec:dos}
\begin{figure}[h]
    \centering
    \begin{minipage}[b]{0.49\textwidth}
        \centering
        \includegraphics[width=\textwidth]{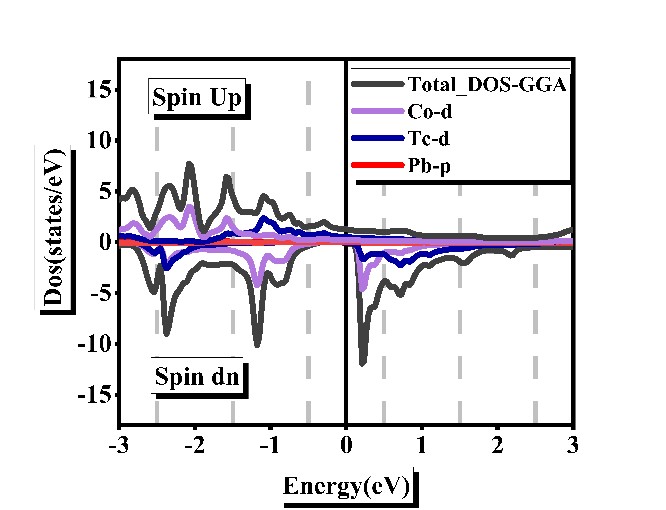}
    \end{minipage}
    \begin{minipage}[b]{0.49\textwidth}
        \centering
        \includegraphics[width=\textwidth]{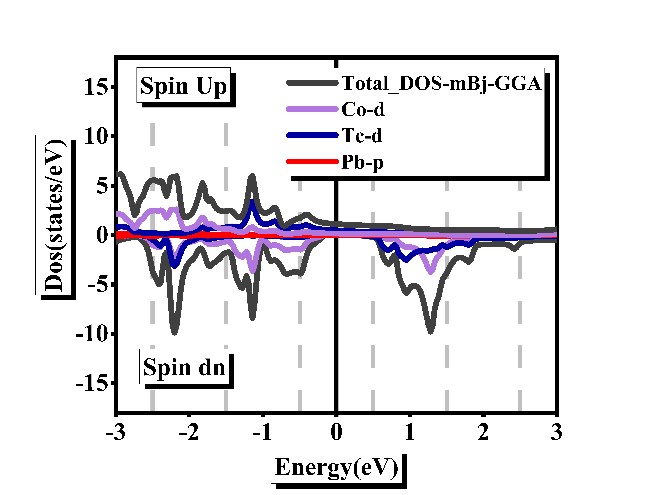}
    \end{minipage}
  \caption{(Colour online) Total (TDOS) and partial (PDOS) density of states for Heusler alloys Co$_2$TcPb calculated using the GGA and GGA-mBJ approximation.}
    \label{fig:6} 
\end{figure}

\begin{figure}[h]
    \centering
    \begin{minipage}[b]{0.49\textwidth}
        \centering
        \includegraphics[width=\textwidth]{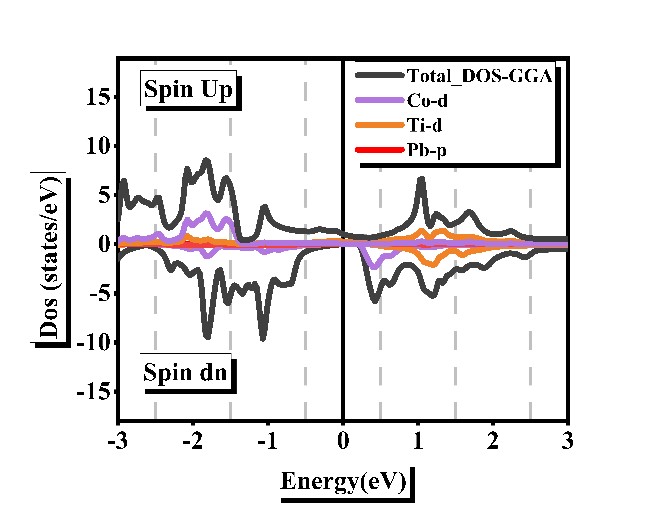}
    \end{minipage}
    \begin{minipage}[b]{0.49\textwidth}
        \centering
        \includegraphics[width=\textwidth]{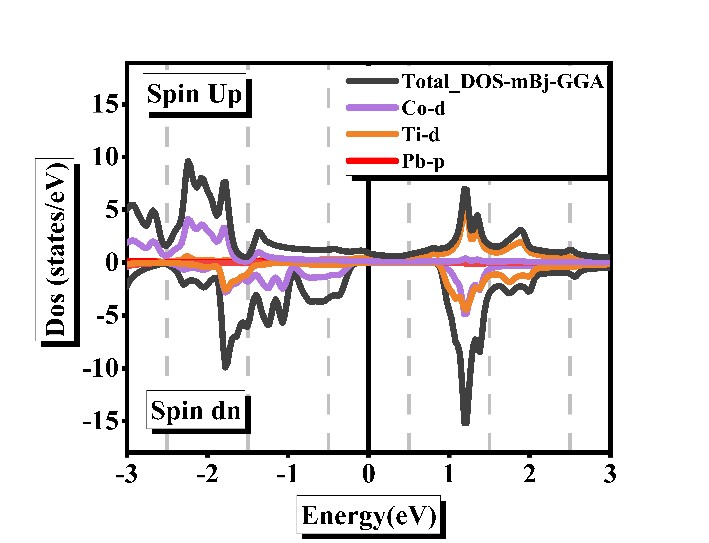}
    \end{minipage}
  \caption{(Colour online) TDOS and PDOS for Heusler alloys Co$_2$TiPb calculated using the GGA and GGA-mBJ approximation.}
    \label{fig:7} 
\end{figure}

\begin{figure}[h]
    \centering
    \begin{minipage}[b]{0.49\textwidth}
        \centering
        \includegraphics[width=\textwidth]{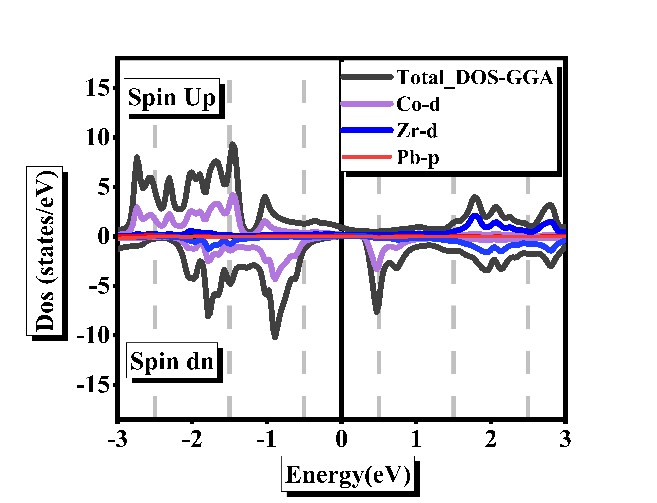}
    \end{minipage}%
    \begin{minipage}[b]{0.49\textwidth}
        \centering
        \includegraphics[width=\textwidth]{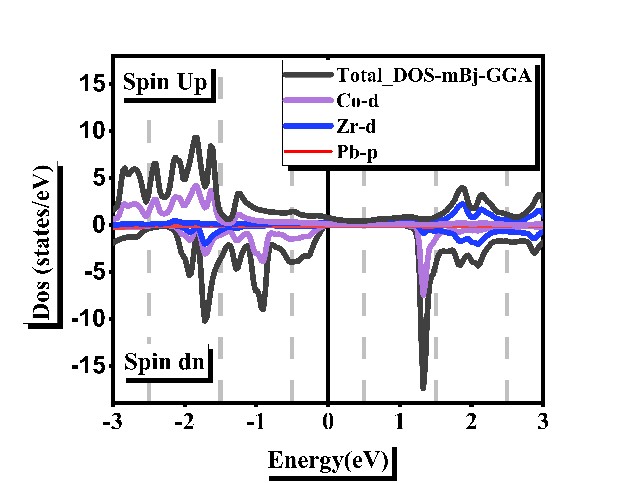}
    \end{minipage}
    \caption{(Colour online) TDOS and PDOS  for the Heusler alloy Co$_2$ZrPb, calculated using the GGA and GGA-mBJ approximations.}
    \label{fig:8}
\end{figure}

\begin{figure}[h]
    \centering
    \begin{minipage}[b]{0.49\textwidth}
        \centering
        \includegraphics[width=\textwidth]{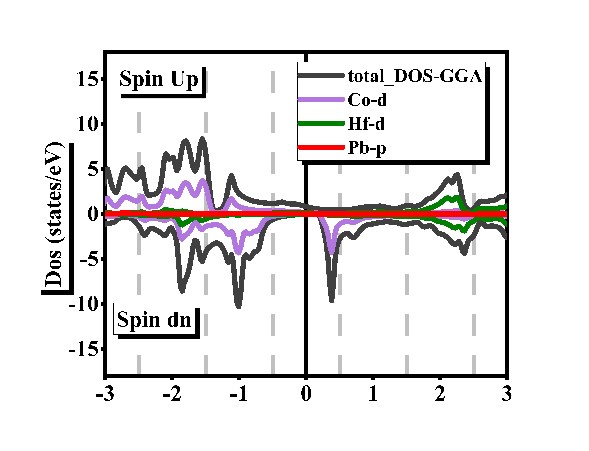}
    \end{minipage}%
    \begin{minipage}[b]{0.49\textwidth}
        \centering
        \includegraphics[width=\textwidth]{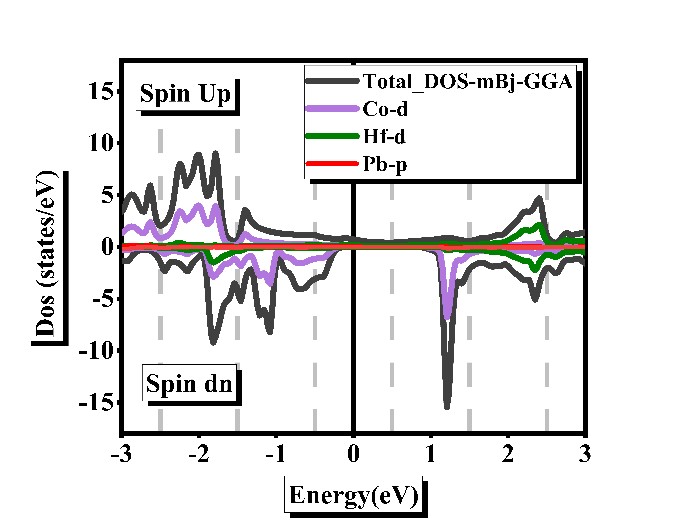}
    \end{minipage}
   \caption{(Colour online) TDOS and PDOS  for the Heusler alloy Co$_2$HfPb, calculated using the GGA and GGA-mBJ approximations.}
    \label{fig:9}
\end{figure}

To fully understand the transition of the conduction band (CB) and valence band (VB), it is necessary to represent the total and projected density of states of the full Heusler X$_2$YZ alloys. In {figures~\ref{fig:6}--\ref{fig:9}}, we exhibit the spin-polarized total and partial density of states to demonstrate the orbital hybridization of Co$_2$YPb (Y = Tc, Ti Zr and Hf) alloys.

The spin-down channel exhibits a semiconductor behavior, through the Fermi level, while the spin-up channel shows metallic properties.
Using the GGA approach, curves in figures~\ref{fig:6}--\ref{fig:9}  show that for both spin directions, the valence band and the conduction band are mainly composed of the Co-$d$ and $d$ states of atoms Tc, Ti, Hf and Zr, respectively, with a minor contribution of the Pb-$p$ states. Co$_2$TiPb, Co$_2$ZrPb, and Co$_2$HfPb compounds exemplified a half-metallic nature whereas Co$_2$TcPb is metallic.

The conduction and valence bands in the spin-down and up channels around the Fermi level are mainly composed of Co($d$) which hybridizes well with the Tc($d$), Ti($d$), Zr($d$) and Hf($d$) compared to Pb($p$) states. In mBJ-GGA approximation, this hybridization is responsible for the half-metallic behavior observed in Co$_2$YPb (Y = Tc, Ti, Zr and Hf).

\newpage

\subsection{Magnetic properties}
\label{subsec:magnetic}
In table~\ref{tab:03}, we present the calculated spin magnetic moments of the constituent atoms in Co$_2$YPb (Y~=~Tc,~Ti,~Zr,~Hf) compounds, as well as the magnetic moment in the interstitial region, labeled as $M_{\rm int}$ which were calculated by GGA and mBj-GGA approximation in the ferromagnetic state of a Cu$_2$MnAl-type structure. In WIEN2k, the magnetic moments are evaluated separately inside the atomic spheres --- referred to as muffin-tin spheres --- and in the interstitial region, which is the space outside these spheres. The values listed under each atom correspond to the integrated spin magnetic moment within their respective muffin-tin spheres. The $M_{\rm int}$ value represents the spin magnetic moment integrated over the interstitial region outside the muffin-tin spheres as obtained in WIEN2k code. The total magnetic moment per formula unit is the sum of the magnetic moments inside the atomic spheres and the interstitial contribution. Galanakis et al. demonstrated that the total spin magnetic moment of full Heusler alloys conforms with the Slater--Pauling rule \cite{Galanakis2002}. 
The rule can be expressed as follows: $\mu_{\text{tot}} = N_V - 24$, where $\mu_{\text{tot}}$ represents the total magnetic moment and $N_V$ represents the total valence electrons per unit cell~\cite{Zheng2012, Birsan2014}. 
We found a total magnetic moment of 2$\mu_{\textrm{B}}$ for Co$_2$TiPb, Co$_2$ZrPb, Co$_2$HfPb and 5$\mu_{\textrm{B}}$ for Co$_2$TcPb.

According to table \ref{tab:03}, in the Co$_2$TcPb compound, both Co and Tc atomic sites exhibit a pronounced spin polarization, with significant local magnetic moments contributing to the overall magnetization. This indicates a strong hybridization and magnetic interaction between Co and Tc atoms. On the other hand, for Co$_2$TiPb, Co$_2$HfPb, and Co$_2$ZrPb, the magnetic moment is predominantly localized on the Co sites, with minimal magnetic contributions from Ti, Hf, and Zr atoms, reflecting their relatively non-magnetic character in these compounds. Importantly, the total magnetic moments per formula unit are integral values, a hallmark signature of half-metallicity in full Heusler alloys. This integer moment arises from a fully spin-polarized electronic structure where one spin channel exhibits a band gap while the other one remains metallic. The results confirm the theoretical prediction of half-metallic ferromagnetism in these Co$_2$YPb compounds, which is of interest for spintronic applications due to the potential for 100\%  spin polarization at the Fermi level.

\begin{table}[!htb]
\caption{The calculated total ($M_{\rm tot}$) and partial ($M_{\rm Co}$, $M_{\rm Y}$, $M_{\rm Pb}$, $M_{\rm int}$) magnetic moments in units of Bohr magneton ($\mu_{\rm B}$) of full-Heusler Co$_2$YPb (Y = Ti, Tc, Zr and Hf) alloys using GGA and mBj-GGA.}
\label{tab:03}
\vspace{2ex}
\begin{center}
\begin{tabular}{llccccc}
\hline
Alloys & Study & $M_{\rm Co}$ & $M_{\rm Y}$ & $M_{\rm Pb}$ & $M_{\rm int}$ & $M_{\rm tot}$ \\
\hline
Co$_2$TcPb & GGA & 1.49 & 1.86 & $-0.01$ & 0.16 & 5.00 \\
           & mBJ-GGA & 1.61 & 1.89 & $-0.05$ & $-0.05$ & 5.00 \\
\hline
Co$_2$TiPb & GGA & 1.10 & $-0.11$ & 0.15 & $-0.10$ & 2.00 \\
           & mBJ-GGA & 1.31 & $-0.37$ & 0.00 & $-0.25$ & 2.00 \\
\hline
Co$_2$ZrPb & GGA & 1.11 & $-0.11$ & 0.01 & $-0.13$ & 2.00 \\
           & mBJ-GGA & 1.28 & $-0.26$ & 0.00 & $-0.32$ & 2.00 \\
\hline
Co$_2$HfPb & GGA & 1.09 & $-0.09$ & $-0.01$ & $-0.11$ & 2.00 \\
           & mBJ-GGA & 1.25 & $-0.21$ & 0.01 & $-0.29$ & 2.00 \\
\hline
\end{tabular}
\end{center}
\end{table}

\subsection{Elastic properties}
\label{subsec:elastic}

The multiplicity of elastic properties of solid materials is very important and necessary because it provides us with information about the stability, ductility, as well as its hardness under the influence of external strong application that changes the shape and size of the material. Since the structure of our compounds is a cube that needs only three elastic constants C$_{11}$, C$_{12}$, C$_{44}$ to meet the following mechanical stability standards (Born criteria):
$C_{11} > 0$, $C_{44} > 0$, $C_{11} > |C_{12}|$, $(C_{11} + 2C_{12}) > 0$. 

It is also worth noting that there are other important mechanical properties of materials, such as bulk modulus ($B$), shear modulus ($G$), Young's modulus ($E$), Poisson's ratio ($\nu$), and anisotropy factor ($A$) that can be obtained through the calculation of elastic constants using the Voigt--Reuss--Hill approxi\-ma\-tion~\cite{Bruhns2014, Hao2008}. 

The bulk modulus ($B$) evaluates the resistance of the material to the volume change, the shear modulus~($G$) determines the resistance of the material to the shape change (plastic deformation), while the Pugh ratio~($B / G$) is used to describe the ductile/brittle properties of materials. If $B/G > 1.75$, the material is more ductile. If $B/G < 1.75$, the material is more brittle \cite{Benatmane2023}. For the case of our compounds Co$_2$YPb (Y = Tc, Ti Zr and Hf), the values of $B/G$ are respectively equal to 7.60, 3.44, 2.25 and 2.00 (see table~\ref{tab:04}). Thus, all compounds are predicted to be ductile.

The Poisson's ratio $\nu$ indicates the material's response in directions perpendicular to the direction of loading. According to the Frantsevich rule, if Poisson’s ratio $\nu > 1/3$, the material is ductile, but if the ratio $\nu < 1/3$, the material is brittle. 
Our calculated values (table~\ref{tab:04}) also suggest ductility for Co$_2$TcPb, Co$_2$TiPb, and Co$_2$ZrPb ($\nu > 1/3$), while Co$_2$HfPb ($\nu = 0.286$) is borderline or potentially brittle based on this criterion alone.

$E$ is the ratio between the stress and strain of a material, which is called the Young's modulus, reflecting the stiffness of the material. A higher Young's modulus $E$ means a more rigid material. The values of Young’s modulus shown in table~\ref{tab:04} are large for our compounds, indicating they are relatively stiff.

The anisotropic ratio ($A$) is a measure of the degree of anisotropy in elastic materials. Anisotropy refers to the property of a material having different mechanical properties in different crystallographic directions. For cubic crystals, it is defined as $A = 2 C_{44} / (C_{11} - C_{12})$. An isotropic material has $A = 1$. For anisotropic materials, $A$ deviates from 1 (either $> 1$ or $< 1$). It is now evident that the shear anisotropy factor ($A$) of the studied compounds (table~\ref{tab:04}) considerably deviates  from 1, therefore, we can conclude that our compounds are anisotropic.

\begin{table}[!htb]
\caption{The calculated elastic constants $C_{ij}$ (GPa), bulk modulus $B$ (GPa),  Reuss ($G_R$) shear moduli~(GPa) and Voigt ($G_V$), Hill shear modulus $G$ (GPa), Pugh’s ratio $B/G$, Young’s modulus $E$ (GPa), Poisson’s ratio $\nu$ and anisotropy factor $A$ at equilibrium volume of full-Heusler Co$_2$YPb (Y = Tc, Ti, Zr and Hf) alloys.}
\label{tab:04}
\vspace{2ex}
\begin{center}
\renewcommand{\arraystretch}{1.1} 
\begin{tabular}{lcccc}
\hline
Property & Co$_2$TcPb & Co$_2$TiPb & Co$_2$ZrPb & Co$_2$HfPb \\
\hline
$C_{11}$ (GPa) & 178.51 & 175.27 & 204.36 & 210.99 \\
$C_{12}$ (GPa) & 162.19 & 140.09 & 118.36 & 117.89 \\
$C_{44}$ (GPa) & 41.77  & 81.72  & 86.01  & 101.60 \\
$B$ (GPa)  & 167.63 & 151.82 & 147.03 & 148.92 \\
$G_R$ (GPa) & 15.77  & 32.29  & 61.42  & 68.97  \\
$G_V$ (GPa) & 28.33  & 55.87  & 68.80  & 79.58  \\
$G$ (GPa)  & 22.05  & 44.08  & 65.11  & 74.23  \\ 
$B/G$      & 7.60   & 3.44   & 2.25   & 2.00   \\
$E$ (GPa)  & 63.38  & 120.58 & 170.22 & 190.96 \\
$\nu$    & 0.43   & 0.36   & 0.311  & 0.286  \\ 
$A$        & 5.11   & 4.64   & 2.00   & 2.18   \\ 
\hline
\end{tabular}
\renewcommand{\arraystretch}{1.0} 
\end{center}
\end{table}

\subsection{Thermoelectric properties}
\label{subsec:thermoelectric}

Electrical energy is essential in our daily lives, both in domestic and industrial settings. Therefore, it is important to find alternatives to non-renewable energy sources, which are causing significant environmental damage. One of the proposed options is the conversion of thermal energy into electrical energy. To calculate the transport properties of the full Heusler Co$_2$YPb (Y = Tc, Ti, Zr and Hf) alloys,  in this study we used the BoltzTraP code \cite{Madsen2006} 
as implemented in WIEN2k package.

With an understanding of the electronic properties of our materials, we decided to calculate the following thermoelectric properties: Seebeck coefficient ($S$), electrical conductivity ($\sigma/\tau$), electronic thermal conductivity ($\kappa_e/\tau$), and the figure of merit ($ZT$) over a temperature range of 50 to 1000~K in order to explore their thermoelectric behavior using a denser $K$-mesh of 120~000~$K$-points. Note that these calculations yield transport coefficients divided by the unknown relaxation time $\tau$. The figure of merit $ZT$ calculation typically assumes a constant relaxation time, which is a simplification.

\subsubsection{Electrical conductivity ($\sigma/\tau$)}
\label{subsubsec:conductivity}

Electrical conductivity is the measure of the capability of a material to conduct electric current, which is typically carried by electrons or ions.
The graph in {figure~\ref{fig:10}a} shows the evolution of electrical conductivity per relaxation time ($\sigma/\tau$) at different temperatures for our compounds. Between 50~K and 300~K, the electrical conductivity remains almost constant, while from 350~K, in the range of values that we took, the electrical conductivity increases continuously within the studied temperature range and reaches its highest calculated value at 1000~K and its values are $6 \times 10^{18}$ S~ms$^{-1}$, $5.6 \times 10^{18}$~S~ms$^{-1}$, $22.3 \times 10^{18}$ S~ms$^{-1}$ and $5.5 \times 10^{18}$ S~ms$^{-1}$ for Co$_2$TcPb, Co$_2$TiPb, Co$_2$ZrPb and Co$_2$HfPb, respectively. The compounds exhibit high electrical conductivity, indicating excellent conduction and low resistivity for an efficient transport of electrical charges with minimal Joule effect losses. We also observe that the value of  conductivity of Co$_2$ZrPb is the highest one, possibly related to details in its band structure near the Fermi level (despite having a large gap in the minority channel).

\subsubsection{Seebeck coefficient ($S$)}
\label{subsubsec:seebeck}

\begin{figure}[htbp]
	\centering
	\begin{minipage}[b]{0.40\textwidth}
		\centering
		\includegraphics[width=\textwidth]{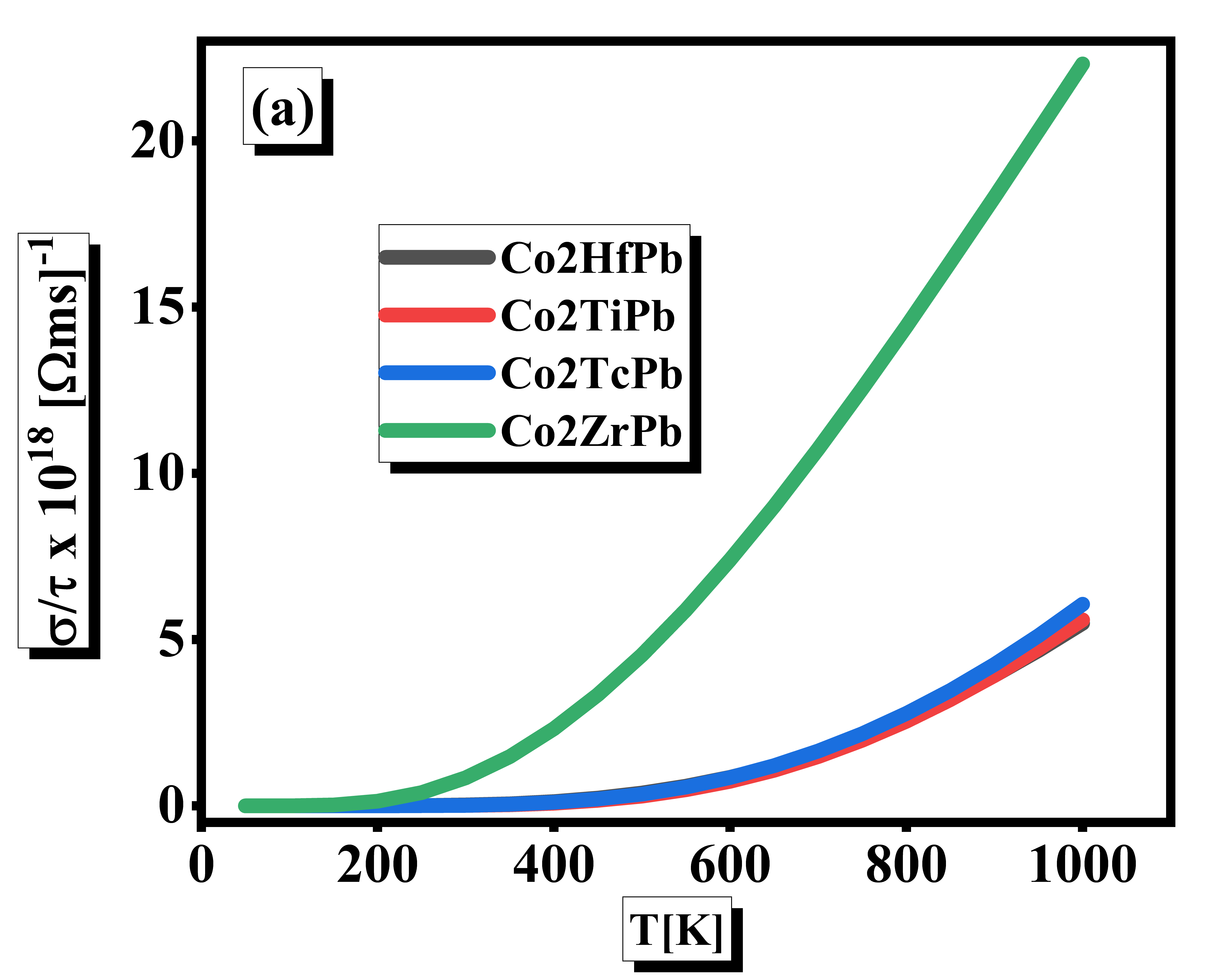}
	\end{minipage}
	\begin{minipage}[b]{0.40\textwidth}
		\centering
		\includegraphics[width=\textwidth]{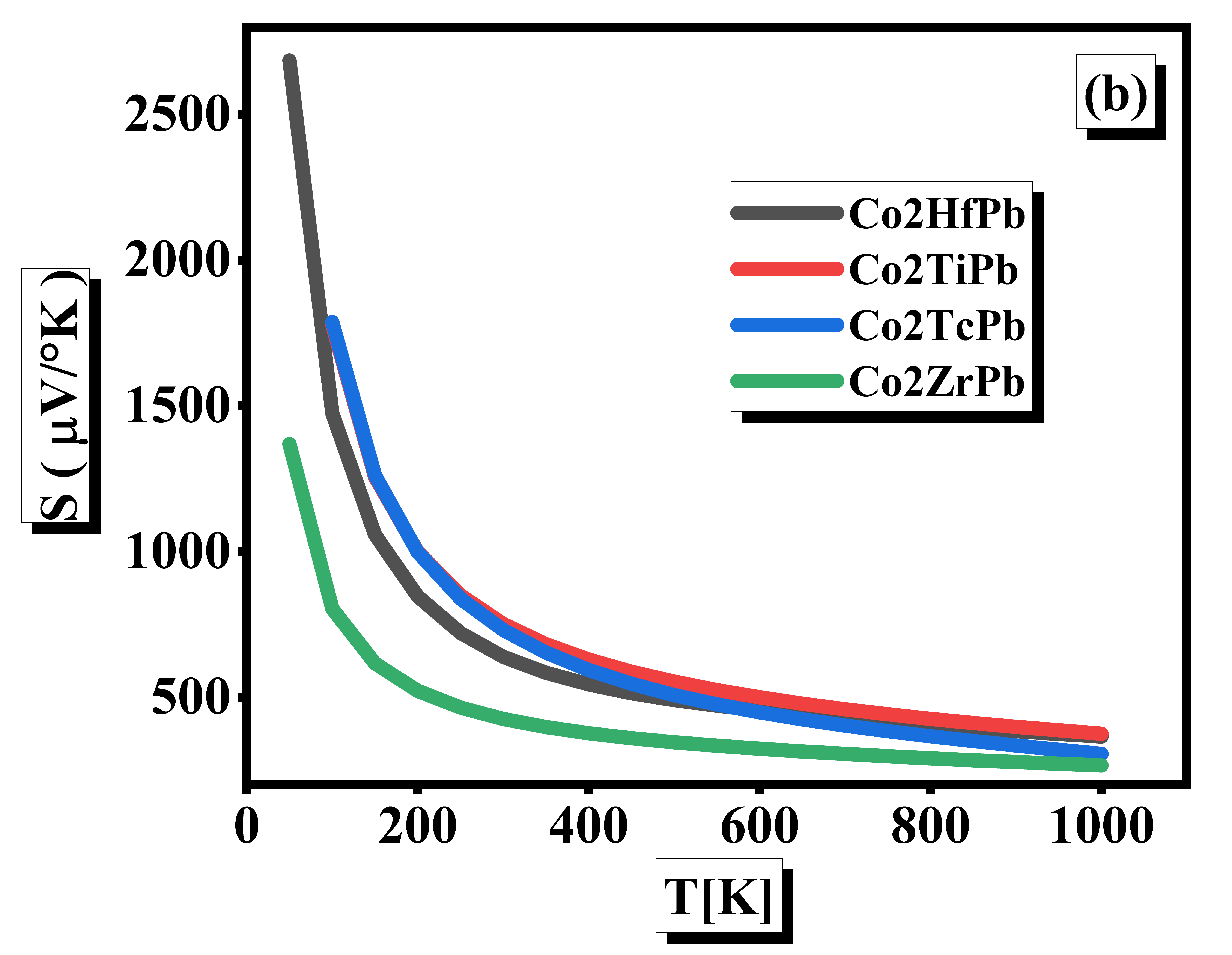}
	\end{minipage}
	\begin{minipage}[b]{0.40\textwidth}
		\centering
		\includegraphics[width=\textwidth]{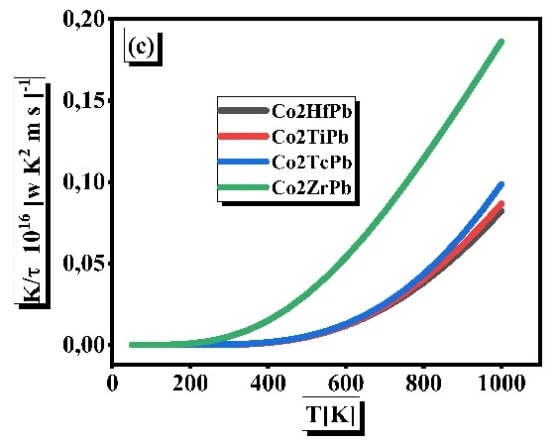}
	\end{minipage}
	\begin{minipage}[b]{0.40\textwidth}
		\centering
		\includegraphics[width=\textwidth]{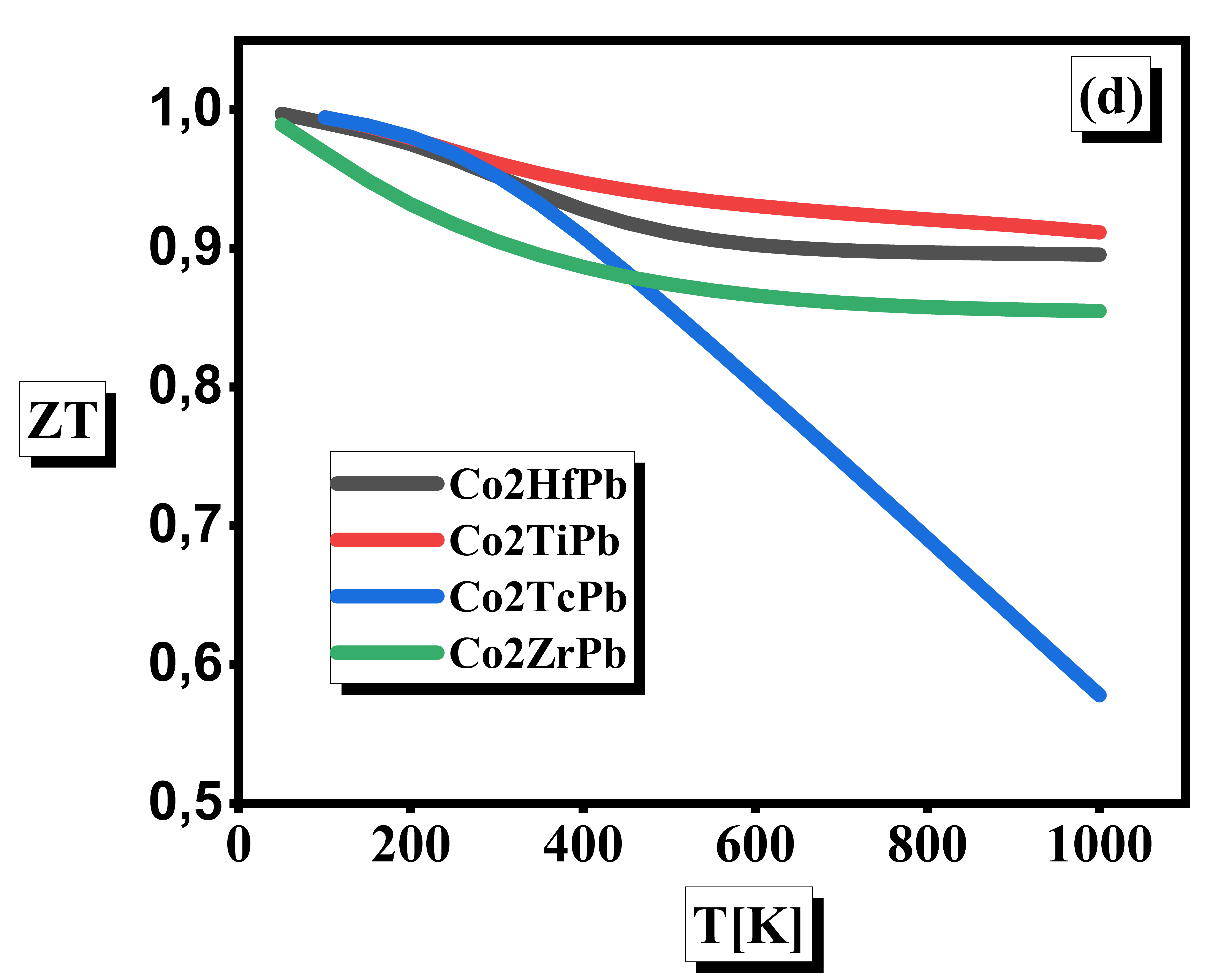}
	\end{minipage}
	\caption{(Colour online) Thermoelectric properties of Co$_2$YPb (Y= Tc, Ti, Zr and Hf) as a function of temperature: (a) electric conductivity ($\sigma/\tau$), (b) Seebeck coefficient ($S$), (c) thermal conductivity ($\kappa/\tau$), and (d) figure of merit ($ZT$).}
	\label{fig:10}
\end{figure}

This is a physical quantity that measures the amplitude of the electrical voltage generated by a material when it is subjected to a temperature gradient. It is defined as the ratio between the electrical potential difference $\Delta V$ induced by a temperature difference $\Delta T$.
It is given as follows $S = \Delta V / \Delta T$. {Figure~\ref{fig:10}b}  presents the changes in the Seebeck coefficient as temperature evolves. 
The positive values indicate $p$-type behavior for all compounds across the temperature range.

At 100~K, the initial values for Co$_2$TcPb, Co$_2$TiPb, Co$_2$ZrPb and Co$_2$HfPb were approximately 1787.8~V~K$^{-1}$, 1778.8~V~K$^{-1}$, 804.9~V~K$^{-1}$ and 1474.8~V~K$^{-1}$, respectively. The Seebeck coefficient for all compounds begins to decrease rapidly with an  increasing temperature, reaching minimum values of 306.7~V~K$^{-1}$ for Co$_2$TcPb, 376.4~V~K$^{-1}$ for Co$_2$TiPb, 267.1~V~K$^{-1}$ for Co$_2$ZrPb and 366.2~V~K$^{-1}$ for Co$_2$HfPb at the temperature of 1000~K. The decrease in the Seebeck coefficient with an  increasing temperature is related to an increase in collisions between charge carriers and phonons, to an increase in electronic thermal conductivity, as well as to a change in diffusion processes and the density of electronic states. This leads to a reduction in the thermal potential difference generated for a given temperature gradient.

\subsubsection{Electronic thermal conductivity ($\kappa_e/\tau$)}
\label{subsubsec:thermalcond}

The electronic thermal conductivity ($\kappa_e/\tau$) of a material refers to its capability to transfer energy in the form of heat via electrons. The total thermal conductivity $\kappa = \kappa_e + \kappa_l$, where $\kappa_l$ represents the contribution from lattice vibrations (phonons) \cite{Boudjeltia2021}. Our calculations only provide the electronic part,~$\kappa_e/\tau$. 

Curves in {figure~\ref{fig:10}c} show the electronic thermal conductivity ($\kappa_e/\tau$) concerning temperature in the range 50--1000~K. The electronic thermal conductivity of Co$_2$TcPb, Co$_2$HfPb and Co$_2$TiPb remains relatively low and constant between 50--350~K, whereas for Co$_2$ZrPb it is constant in the range from 50 to 200~K. Then, we notice a rapid increase in thermal conductivity with temperature. At 1000~K, $\kappa_e/\tau$ reached approximately $0.09 \times 10^{16}$~(W~K$^2$~ms)$^{-1}$ for Co$_2$TcPb, $0.07 \times 10^{16}$~(W~K$^2$~ms)$^{-1}$ for Co$_2$TiPb, $0.19 \times 10^{16}$~(W~K$^2$~ms)$^{-1}$ for Co$_2$ZrPb, and $0.07 \times 10^{16}$~(W~K$^2$~ms)$^{-1}$ for Co$_2$HfPb.
The increase observed in electronic thermal conductivity with temperature follows the Wiedemann--Franz law ($\kappa_e \propto \sigma T$) to some extent, reflecting the increase in electrical conductivity.

\subsubsection{Figure of merit ($ZT$)}
\label{subsubsec:zt}

The figure of merit $ZT$ is used to rate the thermoelectric performance of a material. It is calculated using the formula $ZT = (S^2 \sigma T) / \kappa$. Since we calculated $\sigma/\tau$ and $\kappa_e/\tau$, and lack $\kappa_l$ and $\tau$, the $ZT$ calculated here, most likely using $ZT = [S^2 (\sigma/\tau) T] / (\kappa_e/\tau)$, represents an upper bound assuming $\kappa_l=0$ and requires the relaxation time $\tau$ to cancel out. This provides a relative comparison but not the absolute $ZT$ value.

 {Figure~\ref{fig:10}d} displays how this calculated merit factor $ZT$ changes with temperature. At 100~K, Co$_2$TcPb exhibits a high $ZT$ value approaching 0.99, which rapidly decreases as the temperature increases, reaching a minimum $ZT$ value of approximately 0.57 at 1000~K.
We have also observed that the three other Heusler alloys Co$_2$TiPb, Co$_2$ZrPb and Co$_2$HfPb, exhibit a slight decline in the merit factor $ZT$ as the temperature increases from 100 K to 1000~K. The maximum value of $ZT$ is approximately 0.99, 0.98 and 0.99 at 100~K while the minimum value is 0.91, 0.85 and 0.89 at 1000~K for Co$_2$TiPb, Co$_2$ZrPb and Co$_2$HfPb, respectively. 
Heusler alloys Co$_2$YPb (Y=Ti, Zr, and Hf) exhibit high $ZT$ values (under the calculation assumptions) close to unity, particularly at lower temperatures, making them potentially interesting candidates in the field of thermoelectricity.
The small variation observed in the figure of merit $ZT$ for Co$_2$TiPb, Co$_2$ZrPb, and Co$_2$HfPb with an increasing temperature suggests relatively balanced changes between $S^2\sigma$ and $\kappa_e$ in this range.When these properties change in balance with temperature, the figure of merit can remain relatively stable.

To place the thermoelectric performance of the Co$_2$TiPb, Co$_2$ZrPb, and Co$_2$HfPb compounds into perspective, we compare our results with those of well-established thermoelectric materials. Half-Heusler alloys such as CoTiSb and ZrNiSn have been extensively investigated and are known to exhibit a promising thermoelectric behavior in the moderate temperature range (300--800~K), with typical figure of merit ($ZT$) values approaching or exceeding 1 under optimized conditions \cite{Chen2013}. These materials are already considered viable for integration into waste heat recovery systems in automotive and industrial environments.

In addition, full Heusler compounds have drawn an increasing attention due to their tunable electronic and magnetic properties, mechanical stability, and compatibility with high-temperature thermoelectric applications \cite{Graf2011}. Recent reviews have also emphasized the relevance of Heusler and half-Heusler alloys for embedded systems in the automotive sector, especially for energy harvesting applications and thermal management in harsh conditions \cite{Schierning2021}.

Compared to these well known systems, the Co$_2$YPb (Y = Ti, Zr, Hf) compounds studied here display competitive Seebeck coefficients and power factors. Their performance, combined with the possibility of tailoring their band structure and magnetic behavior through elemental substitution, positions them as promising candidates for next-generation thermoelectric devices, particularly in embedded automotive applications.


\section{Conclusion}
\label{sec:conclusion}

In this paper, we performed first-principles calculations to investigate the structural, electronic, magnetic, elastic and thermoelectric properties of full Heusler Co$_2$YPb alloys (Y = Tc, Ti, Zr and Hf).
We determined that these compounds are more stable in the Cu$_2$MnAl-type structure in the ferromagnetic state compared to the Hg$_2$CuTi-type structure.
Electronic band structure and calculations of density of states  reveal that all these compounds have a half-metallic property while using mBj-GGA.
The total magnetic moment of all the full-Heusler alloys studied adopts an integer value, in accordance with the Slater--Pauling rule which confirms the half metallicity of the compounds. This electronic structure: metallic in one spin channel and semiconducting in the other, leads to full spin polarization at the Fermi level, opening prospects for spintronic applications. The analysis of the elastic properties indicates that the Co$_2$YPb (Y = Tc, Ti, Zr, and Hf) full Heusler alloys are mechanically stable in their cubic phase and exhibit a ductile character along with noticeable elastic anisotropy. Moreover, their moderate stiffness suggests a balanced mechanical response, which may be advantageous for integration into functional devices. Thermoelectric properties were also studied by varying the Seebeck coefficient, electrical conductivity, total thermal conductivity and figure of merit ($ZT$) as a function of temperature. The Full-Heusler Co$_2$YPb (Y = Tc, Ti, Zr, Hf) alloys are $p$-type semiconductors, since their Seebeck coefficients are positive over the entire temperature range studied. Three alloys show high $ZT$ values, close to unity, at medium temperatures.

The promising thermoelectric behavior of Co$_2$TiPb, Co$_2$ZrPb, and Co$_2$HfPb, particularly their high and stable $ZT$ values around 1000 K, make them strong candidates for integration into embedded thermoelectric generator (TEG) systems in the automotive sector. TEGs are used to recover waste heat from engine exhausts or other high-temperature components and convert it into electrical energy, which can power the onboard electronics or reduce the fuel consumption. For such applications, materials must maintain an efficient energy conversion at high temperatures, demonstrate mechanical and chemical stability, and be compatible with standard fabrication techniques. The full-Heusler structure of our compounds is known for its robustness and resistance to degradation under thermal stress. Additionally, their performance in the 0--1000~K range aligns with typical automotive operating conditions. These characteristics support the practical feasibility of implementing these materials in next-generation embedded automotive systems.

\newpage

\ukrainianpart

\title{Першопринципне дослідження електронних, магнітних та термоелектричних властивостей повних сполук Гейслера  Co$_2$YPb (Y = Tc, Ti, Zr та Hf): застосування до вбудованих автомобільних систем}
\author{Н. Саїді, А. Аббад, В. Бенстаалі, К. Бане}
\address{
	Лабораторія технології та властивостей твердих тіл, факультет наук і технологій, BP227, Університет Абдельхаміда Ібн Бадіса, 27000 Мостаганем, Алжир
}

\makeukrtitle

\begin{abstract}
	\tolerance=3000%
	У рамках методу функціоналу густини (DFT) проведено теоретичні дослідження структурних, електронних, магнітних, пружних та термоелектричних властивостей повних сплавів Гейслера Co$_2$YPb (Y = Tc, Ti, Zr та Hf). Обмінний потенціал та кореляції розглядаються за допомогою двох наближень: узагальненого градієнтного наближення (GGA) та GGA, доповненого модифікованим Траном-Блахою наближенням Бекке-Джонсона (mBj-GGA), яке забезпечує точніший опис ширини забороненої зони. Електронні та магнітні властивості показують, що повні сплави Гейслера Co$_2$YPb (з Y = Tc, Ti, Zr та Hf) проявляють напівметалічну феромагнітну поведінку. Крім того, пружні властивості свідчать, що Co$_2$YPb є механічно стійкими та мають пластичні характеристики. Повні сплави Гейслера $p$-типу мають додатні коефіцієнти Зеебека та високі значення коефіцієнта якості, що вказує на хороші термоелектричні характеристики з точки зору електро- та теплопровідності. Це дозволяє стверджувати, що ці сполуки є дуже цікавими для покращення продуктивності вбудованих автомобільних систем, а також можуть бути використані в спінтронних пристроях.
	\keywords термоелектрика, напівметали, магнітні сполуки, вбудовані системи, mBJ-GGA
	
\end{abstract}

\lastpage

\begin{thebibliography}{99}
\bibitem{Wollmann2017} 
Wollmann L., Nayak A. K., Parkin S. S. P., Felser C., 
Annu. Rev. Mater. Res., 2017, {\bf 47}, No.~1, 247--270, \\ \doi{10.1146/annurev-matsci-070616-123928}.
\bibitem{Bradley1934} 
Bradley A. J., Rodgers J. W., 
Proc. R. Soc. London, Ser. A, 1934,  {\bf 144}, No.~852, 340--359, \doi{10.1098/rspa.1934.0053}.  
\bibitem{Chadov2011} 
Chadov S., Graf T., Chadova K., Dai X., Gasper F., Fecher G. H., Felser C., 
Phys. Rev. Lett., 2011, {\bf 107}, No.~4, 047202, \doi{10.1103/PhysRevLett.107.047202}.
\bibitem{Hara2006} 
Hara M., Shibata J., Kimura T., Otani Y., 
Appl. Phys. Lett., 2006, {\bf 88}, 082501, \doi{10.1063/1.2177358}.
\bibitem{Wolf2001} 
Wolf S. A., Awschalom D. D., Buhrman R. A., Daughton J. M., von Moln\'{a}r S., Roukes~M.~L., Chtchelkanova~A.~Y., Tregger D. M., 
Science, 2001, {\bf 294}, No. 5546, 1488--1495,  \doi{10.1126/science.1065389}.
\bibitem{DeGroot1983} 
de Groot R. A., Mueller F. M., van Engen P. G., Buschow K. H. J., 
Phys. Rev. Lett., 1983, {\bf 50}, No.~25, 2024,\\ \doi{10.1103/PhysRevLett.50.2024}.
\bibitem{Kawasaki2022} 
Kawasaki J. K., Chatterjee S., Canfield P. C., 
MRS Bull., 2022, {\bf 47}, No.~6, 555--558, \doi{10.1557/s43577-022-00355-w}.
\bibitem{Ouardi2010} 
Ouardi S., Fecher G. H., Balke B., Kozina X., Stryganyuk G., Felser C., Lowitzer S., K\"odderitzsch D., Ebert H., Ikenaga E., 
Phys. Rev. B, 2010, {\bf 82}, No.~8, 085108, \doi{10.1103/PhysRevB.82.085108}.
\bibitem{Kubler1983} 
Kübler J., William A. R., Sommers C. B., 
Phys. Rev. B, 1983, {\bf 28}, No.~4, 1745, \doi{10.1103/PhysRevB.28.1745}.
\bibitem{Junxiang2023} 
Junxiang Y., Kumar P., Cabero-Piris M., Aarts J., 
Phys. Rev. Mater., 2023,  {\bf 7}, No.~10, 104408, \\ \doi{10.1103/PhysRevMaterials.7.104408}.
\bibitem{Kostenko2018} 
Kostenko M. G., Lukoyanov M. V., Shreder E. I., 
JETP Lett., 2018, {\bf 107}, 126--128,\\ \doi{10.1134/S002136401802008X}.
\bibitem{Ivanshin2009} 
Ivanshin V. A., Litvinova T. O., Sukhanov A. A., Sokolov D. A., Aronson M. C., 
JETP Lett., 2009, {\bf 90}, 116--119,\\ \doi{10.1134/S0021364009140070}.
\bibitem{Marchenkov2023} 
Marchenkov V. V., Irkhin V. Yu., Marchenkova E. B., Semiannikova A. A., Korenistov P. S., 
Phys. Lett. A, 2023, {\bf 471}, 128803, \doi{10.1016/j.physleta.2023.128803}.
\bibitem{Graf2009} 
Graf T., Casper F., Winterlik J., Balke B., Fecher G. H., Felser C., 
Z. Anorg. Allg. Chem., 2009, {\bf 635}, No.~6--7, 976--981, \doi{10.1002/zaac.200900036}.
\bibitem{Salaheldeen2022} 
Salaheldeen M., Garcia-Gomez  A., Ipatov M., Corte-Leon P., Zhukova V., Blanco J. M., Zhukov A., 
Chemosensors, 2022, {\bf 10}, No.~6, 225, \doi{10.3390/chemosensors10060225}.
\bibitem{Benatmane2020} 
Benatmane S., Cherid S., 
JETP Lett., 2020, {\bf 111}, 694--702, \doi{10.1134/S0021364020120012}.
\bibitem{Zitouni2020} 
Zitouni A., Remil G., Bouadjemi B., Benstaali W., Lantri T., Matougui M., Houari M., Aziz Z., Bentata S., 
JETP Lett., 2020, {\bf 112}, 290--298, \doi{10.1134/S0021364020170026}.
\bibitem{Perdew1998} 
Perdew J. P., Burke K., Ernzerhof M., 
Phys. Rev. Lett., 1998, {\bf 80}, No.~4, 891, \doi{10.1103/PhysRevLett.80.891}.
\bibitem{Tran2009} 
Tran F., Blaha P., 
Phys. Rev. Lett., 2009, {\bf 102}, No.~22, 226401, \doi{10.1103/PhysRevLett.102.226401}.
\bibitem{Tran2007} 
Tran F., Blaha P., Schwarz K., 
J. Phys.: Condens. Matter, 2007, {\bf 19}, No.~19, 196208, \doi{10.1088/0953-8984/19/19/196208}.
\bibitem{Pagare2014} 
Pagare G., Chouhan S. S., Soni P., Sanyal S. P., Rajagopalan M., 
Comput. Mater. Sci., 2010, {\bf 50}, 538--544,\\ \doi{10.1016/j.commatsci.2010.09.016}.
\bibitem{Monkhorst1976} 
Monkhorst H. J., Pack J. D., 
Phys. Rev. B, 1976, {\bf 13}, No.~12, 5188, \doi{10.1103/PhysRevB.13.5188}.
\bibitem{Houari2019} 
Houari M., Bouadjemi B., Haid S., Matougui M., Lantri T., Aziz Z., Bentata S., Bouhafs B., Indian J. Phys., 2020, {\bf 94}, 455, \doi{10.1007/s12648-019-01480-0}.
\bibitem{Mentefa2021} 
Mentefa A., Boufadi F. Z., Ameri M., Gaid F. O., Bellagoun L., Odeh A. A., Al-Douri Y., 
J.~Supercond.~Novel~Magn., 2021, {\bf 34}, 269--283, \doi{10.1007/s10948-020-05741-6}.
\bibitem{Murnaghan1944} 
Murnaghan F. D., 
Proc. Natl. Acad. Sci. U.S.A., 1944, {\bf 30}, No.~9, 244--247, \doi{10.1073/pnas.30.9.244}.
\bibitem{Karimian2015} 
Karimian N., Ahmadian F., 
Solid State Commun., 2015, {\bf 223}, 60--66, \doi{10.1016/j.ssc.2015.09.005}.
\bibitem{Idriss2020} 
Idriss S., Labrim H., Ziti S., Bahmad L., 
Appl. Phys. A, 2020, {\bf 126}, 190, \doi{10.1007/s00339-020-3354-6}.
\bibitem{Houari2024} 
Houari M., Mesbah S., Lantri T., Bouadjemi B., Boucherdoud A., Khatar A., Akham A., Haid S., Achour B., Bentata S., Matougui M., 
 J. Mol. Model., 2024,  {\bf 30}, 110, \doi{10.1007/s00894-024-05903-6}.
\bibitem{Galanakis2002} 
Galanakis I., Dederichs P. H., Papanikolaou N.,
Phys. Rev. B, 2002, {\bf 66}, No.~17, 174429,\\ \doi{10.1103/PhysRevB.66.174429}.
\bibitem{Zheng2012} 
Zheng N., Jin Y., 
J. Magn. Magn. Mater., 2012, {\bf 324}, No.~19, 3099--3104, \doi{10.1016/j.jmmm.2012.05.009}.
\bibitem{Birsan2014} 
Birsan A., 
Curr. Appl. Phys., 2014, {\bf 14}, No.~11, 1434--1436, \doi{10.1016/j.cap.2014.08.009}.
\bibitem{Bechmann1958} 
Bechmann R., 
Phys. Rev., 1958, {\bf 110}, No.~5, 1060,  \doi{10.1103/PhysRev.110.1060}.
\bibitem{Bruhns2014} 
Bruhns O. T., 
J. Appl. Math. Mech., 2014, {\bf 94}, No.~3, 187--202, \doi{10.1002/zamm.201300243}.
\bibitem{Hao2008} 
Hao Y. J., Zhang L., Chen X. R., Li Y. H., He H. L., 
J. Phys.: Condens. Matter, 2008, {\bf 20}, No.~23, 235230,\\ \doi{10.1088/0953-8984/20/23/235230}.
\bibitem{Benatmane2023} 
Benatmane S., Affane M., Bouali Y., Bouadjemi B., Cherid S., Benstaali W., 
Rev. Mex. Fis., 2023, {\bf 69}, No.~1, 011003, \doi{10.31349/RevMexFis.69.011003}.
\bibitem{Madsen2006} 
Madsen G. K. H., Singh D. J., 
 Comput. Phys. Commun., 2006, {\bf 175}, No.~1, 67--71, \doi{10.1016/j.cpc.2006.03.007}.
\bibitem{Boudjeltia2021} 
Boudjeltia M. A., Aziz Z., Terkhi S., Bennani M. A., Khandy S. A., Bouadjemi B., Benidris M., Bentata S., 
Mod. Phys. Lett. B, 2021, {\bf 35}, No.~23, 2150400, \doi{10.1142/S0217984921504005}.
\bibitem{Chen2013}
Chen S., Ren Z., 
Mater. Today, 2013, {\bf 16}, No.~10,  387, \doi{10.1016/j.mattod.2013.09.015}.
\bibitem{Graf2011}
Graf T., Felser C., Parkin S. S. P., 
Prog. Solid State Chem., 2011, {\bf 39}, No. 1, 1,\\ \doi{10.1016/j.progsolidstchem.2011.02.001}.
\bibitem{Schierning2021}
 Albaladejo-Siguan M., Baird E. C., Becker-Koch D., Li Y., Rogach A. L., Vaynzof Y.,   
Adv. Energy Mater., 2021, {\bf 11},  2003457, \doi{10.1002/aenm.202003457}.

\end{thebibliography}
\end{document}